\documentclass[aps,prb,showpacs,twocolumn,floats,psfig]{revtex4}
\usepackage{amssymb}
\usepackage{amsbsy}
\usepackage{amsmath}
\usepackage{epsfig}
\newcommand\bea{\begin{eqnarray}}
\newcommand\eea{\end{eqnarray}}
\newcommand\beq{\begin{equation}}
\newcommand\eeq{\end{equation}}

\newcommand{\non}{\nonumber}
\newcommand{\bib}{\bibitem}
\newcommand{\de}{\delta}
\newcommand{\si}{\sigma}
\newcommand{\ta}{\theta}
\newcommand{\da}{\dagger}
\newcommand{\la}{\langle}
\newcommand{\ra}{\rangle}
\newcommand{\vk}{\vec k}
\newcommand{\vn}{\vec n}
\newcommand{\vcr}{\vec r}

\begin{document}

\title{Quench dynamics and defect production in the Kitaev and extended
Kitaev models}

\author{Shreyoshi Mondal$^1$, Diptiman Sen$^2$ and K. Sengupta$^1$}
\affiliation{$^1$ TCMP division, Saha Institute of Nuclear Physics,
1/AF Bidhannagar, Kolkata 700 064, India \\ $^2$ Center for High Energy
Physics, Indian Institute of Science, Bangalore, 560 012, India}

\date{\today}

\begin{abstract}
We study quench dynamics and defect production in the Kitaev and the
extended Kitaev models. For the Kitaev model in one dimension, we
show that in the limit of slow quench rate, the defect density $n
\sim 1/\sqrt{\tau}$ where $1/\tau$ is the quench rate. We also
compute the defect correlation function by providing an exact
calculation of all independent non-zero spin correlation functions
of the model. In two dimensions, where the quench dynamics takes the
system across a critical line, we elaborate on the results of
earlier work [K. Sengupta, D. Sen and S. Mondal, Phys. Rev. Lett.
{\bf 100}, 077204 (2008).] to discuss the unconventional scaling of
the defect density with the quench rate. In this context, we outline
a general proof that for a $d$ dimensional quantum model, where the
quench takes the system through a $d-m$ dimensional gapless
(critical) surface characterized by correlation length exponent
$\nu$ and dynamical critical exponent $z$, the defect density $n
\sim 1/\tau^{m\nu/(z\nu +1)}$. We also discuss the variation of the
shape and the spatial extent of the defect correlation function with
the change of both the rate of quench and the model parameters and
compute the entropy generated during such a quench process. Finally,
we study the defect scaling law, entropy generation and defect
correlation function of the two-dimensional extended Kitaev model.
\end{abstract}


\maketitle

\section{Introduction}

Quantum phase transitions involve a fundamental change in the
symmetry of the ground state of a quantum system. Such a transition
usually takes place due to the variation of some parameter $\lambda$
in the Hamiltonian of the system and is necessarily accompanied by
diverging length and time scales \cite{sachdev}. A direct
consequence of such a diverging time scale is that a quantum system
fails to be in the adiabatic limit when it is sufficiently close to
the quantum critical point. Thus a time evolution of the parameter
$\lambda$ at a finite rate $1/\tau$, which takes such a system
across a quantum critical point located at $\lambda=\lambda_c$,
leads to failure of the system to follow the instantaneous ground
state in a finite region around $\lambda_c$. As a result, the state
of the system after such a time evolution does not conform to the
ground state of its final Hamiltonian leading to the production of
defects \cite{kz1,damski1}. It is well known that for a slow quench, the
density of these defects $n$ depends on the quench time $\tau$
according to $n \sim 1/\tau^{d\nu/(\nu z+1)}$, where $\nu$ and $z$
are the correlation length and the dynamical critical exponents
characterizing the critical point \cite{anatoly1,anatoly2,comment1}.
A theoretical study of such a quench dynamics requires a knowledge
of the excited states of the system. As a result, early studies of
the quench problem are mostly restricted to quantum phase
transitions in exactly solvable models such as the one-dimensional
(1D) Ising model in a transverse field \cite{ks1,dziar1,cardy}, the
infinite range ferromagnetic Ising model \cite{das}, the 1D XY
model \cite{levitov,sen1}, and quantum spin chains
\cite{damski2,caneva,zurek}. On the experimental side, trapped ultracold
atoms in optical lattices provide possibilities of realization of
many of the above-mentioned systems \cite{bloch}. Experimental
studies of defect production due to quenching of the magnetic field
in a spin-one Bose condensate has also been undertaken \cite{sadler}.

Recently, Kitaev proposed a 2D spin-1/2 model on a honeycomb lattice
with a Hamiltonian \cite{kitaev1}
\bea H_{1} &=& \sum_{j+l={\rm even}} ~(~ J_1 \si_{j,l}^x \si_{j+1,l}^x ~+~
J_2 \si_{j-1,l}^y \si_{j,l}^y \non \\
& & ~~~~~~~~~~~~~~+ J_3 \si_{j,l}^z \si_{j,l+1}^z ~), \label{kham1}
\eea where $j$ and $l$ denote the column and row indices of the
honeycomb lattice. This model has several interesting features which
led to a plethora of theoretical works on it
\cite{feng,baskaran,vidal}. For example, it provides a rare example
where a 2D model can be exactly solved using a Jordan-Wigner transformation
\cite{kitaev1,feng,nussinov1,nussinov2}. Further, when $J_3=0$, the
model provides an example of an 1D spin model which supports a topological
quantum phase transition with the critical point at $J_1=J_2$
\cite{feng}. Moreover, in $d=2$, the model supports a gapless phase
for $|J_1-J_2|\le J_3 \le J_1+J_2$ which has a possible connection
to a spin liquid state and demonstrates fermion fractionalization at
all energy scales \cite{baskaran}. Finally, it has been shown in
Ref.\ \onlinecite{kitaev1} that the presence of magnetic field, which
induces a gap in the 2D gapless phase, leads to non-Abelian statistics of
the low-lying excitations of the model; these excitations can be viewed as
robust qubits in a quantum computer \cite{kitaev2}. An extended version
of this model has also been suggested in Ref.\ \onlinecite{lee} which has
the Hamiltonian
\bea H_2 &=& J_4 ~\Bigg [ \sum_{j+l={\rm odd}}\sigma_{j,l}^y \sigma_{j+1,l}^z
\sigma_{j+2,l}^x \non \\
&& ~~~ + \sum_{j+l={\rm even}}\sigma_{j,l}^x\sigma_{j+1,l}^z\sigma_{j+2,l}^y
\Bigg] ~+~ H_1. \label{extkit1} \eea

The quench dynamics of the 2D Kitaev model has been studied very
recently in Ref.\ \onlinecite{sms}. It has been shown that for this
model, quenching $J_3$ takes the system through a critical line
instead of critical point which leads to unconventional variation of
the defect density as a function of the quench rate. In this
context, it has also been shown that for a general $d$-dimensional
model, where the quench take the system through a $d-m$ dimensional
hypersurface characterized by the correlation length exponent $\nu$
and dynamical critical exponent $z$, the defect density obeys $n_d
\sim 1/\tau^{m \nu/(z \nu +1)}$. The Kitaev model provides a
concrete example of such a quench for $d=2$ and $m=1$. The defect
correlation function for such a quench has also been computed in
Ref.\ \onlinecite{sms}.

In this work, we extend and elaborate on the results of Ref.\
\onlinecite{sms} and study the quench dynamics of the Kitaev model
both in $d=1$ and $d=2$ and the extended Kitaev model in $d=2$. The
main results that we have obtained are the following. First, we show
that in 1D ($J_3=0$), where quenching $J_1$ takes the system across
the topological quantum critical point located at $J_1=J_2$, the
density of defects produced due to the quench scales as
$1/\sqrt{\tau}$ in the limit of slow quench (large $\tau$). We also
identify and compute all independent non-zero spin-spin correlation
functions and use them to elucidate the spatial extent of the defect
correlation function. Second, we outline a general proof of the
result reported in Ref.\ \onlinecite{sms} that for a $d$ dimensional
quantum model, where the quench take the system through a $d-m$
dimensional hypersurface characterized by the correlation length
exponent $\nu$ and dynamical critical exponent $z$, the defect
density obeys $n_d \sim 1/\tau^{m \nu/(z \nu +1)}$. Third, we
elaborate on the variation of shape and size of the defect
correlation function for the 2D Kitaev model with the quench rate
and the model parameters. Fourth, we compute the entropy generated
due to such a quench and discuss its dependence on the model
parameters and the quench rate. Finally, we study the defect scaling
law, entropy generation and defect correlation function of the 2D
extended Kitaev model described by $H_2$.

\begin{figure}
\rotatebox{0}{\includegraphics*[width=\linewidth]{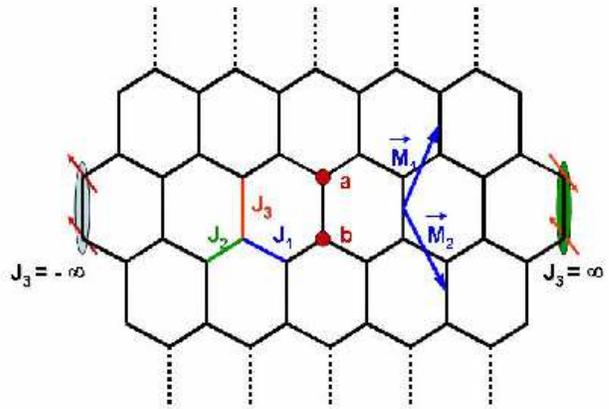}}
\caption{Schematic representation of the Kitaev model on a honeycomb
lattice showings the bonds $J_1$, $J_2$ and $J_3$. Schematic
pictures of the ground states, which correspond to pairs of spins on
vertical bonds locked parallel (antiparallel) to each other in the
limit of large negative (positive) $J_3$, are shown at one bond on
the left (right) edge respectively. ${\vec M}_1$ and ${\vec M}_2$
are spanning vectors of the lattice, and $a$ and $b$ represent
inequivalent sites.} \label{fig0} \end{figure}

The organization of the paper is as follows. In Sec.\ \ref{1da}, we
analyze the quench dynamics of the Kitaev model in 1D and obtain the
quench rate dependence of the defect density. This is followed by
Sec.\ \ref{1db}, where we compute the 1D correlation functions and
use them to discuss the nature of the defect correlation function.
Next, in Sec.\ \ref{2da}, we obtain the quench rate dependence of
the defect density in 2D. The computation of the defect correlation
function is detailed in Sec.\ \ref{2db} and the entropy generated
during the quench process is computed in Sec.\ \ref{ent}. This is
followed by the study of quench dynamics of the extended Kitaev
model in Sec.\ \ref{ekit}. Finally, we conclude in Sec.\ \ref{concl}.

\section{Quench in 1D}
\label{1d}

\subsection{Defect density}
\label{1da}

For $J_3=0$, the Kitaev model represents a spin-1/2 model in 1D with the
Hamiltonian
\bea H_{\rm 1D}&=& \sum_{n} \left(J_1 \si_{2n}^x \si_{2n+1}^x + J_2
\si_{2n-1}^y \si_{2n}^y \right) \label{hamkit1d1}, \eea
where $n$ denotes site indices of a one dimensional chain with $N$ sites (we
will assume $N$ is a multiple of 4). The lattice spacing $a$ and the
Planck constant $\hbar$ will be set equal to $1$ in the rest of this work.
The Hamiltonian in Eq. (\ref{hamkit1d1}) can be exactly
diagonalized using a standard Jordan-Wigner transformation \cite{kogut}
\bea a_n &=& \left( \prod_{j=-\infty}^{2n-1} ~\si_j^z \right) ~\si_{2n}^y,
\non \\
b_n &=& \left( \prod_{j=-\infty}^{2n} ~\si_j^z \right) ~\si_{2n+1}^x ,
\label{maj1d} \eea
where $a_n$ and $b_n$ are independent Majorana fermions at site $n$. They
satisfy relations such as $a_n^\da = a_n$, $\{ a_m , a_n \} = 2 \de_{m,n}$
and $\{ a_m , b_n \} = 0$. The label $n$ for $a_n$ and $b_n$ go over $N/2$
values since that is the number of unit cells. In terms of these operators,
$H_{\rm 1D}$ can be written as
\bea H_{\rm 1D} &=& i ~\sum_n ~[~ J_1 ~b_n a_n ~+~ J_2 ~b_n a_{n+1} ~] \non \\
&=& 2i ~\sum_{k=0}^\pi ~[~ b_k^\da a_k ~(J_1 + J_2 e^{ik}) \non \\
& & ~~~~~~~~~~~+~ a_k^\da b_k ~(- J_1 - J_2 e^{-ik}) ~], \label{hamkit1d2} \eea
where the Majorana fermion creation and destruction operators $a_k^\da$ and
$a_k$ are Fourier components of the $a_n$'s,
\bea a_n ~=~ \sqrt{\frac{4}{N}} ~\sum_{k=0}^\pi ~[~ a_k ~e^{ikn} ~+~ a_k^\da ~
e^{-ikn} ~]. \label{fourier1} \eea
The sum over $k$ in Eq. (\ref{fourier1}) only goes over half the Brillouin zone
because $a_n$ describes a Majorana fermion; the number of modes lying in the
range $0 \le k \le \pi$ is $N/4$. [There is a small correction that one
has to make in Eq. (\ref{fourier1}) for the modes with $k=0$ and $\pi$ for
which there is no distinction between $k$ and $-k$; these two modes
should have a coefficient of $\sqrt{2/N}$ instead of $\sqrt{4/N}$.
However, we will ignore this correction here because we will be interested
in the $N \to \infty$ limit, and we will change from a sum over $k$
to an integral over $k$.] The operators $a_k$ and $a_k^\da$ satisfy the
anticommutation relations $\{ a_k , a_{k'}^\da \} = \de_{kk'}$ and $\{ a_k ,
a_{k'} \} = 0$. One can now define a two component fermionic creation
operator $\psi_k = (a_k ~b_k)$, so that $H_{\rm 1D}$ can be written as
\bea H_{\rm 1D} &=& \sum_{k=0}^\pi \psi_k^\da ~H_k \psi_k , \non \\
{\rm where} ~~H_k &=& 2i ~\left( \begin{array}{cc} 0 & -J_1 - J_2 e^{-ik} \\
J_1 + J_2 e^{ik} & 0 \end{array} \right). \label{kitham1d3} \eea
{}From Eq.\ (\ref{kitham1d3}), we find that $H_{\rm 1D}$ can be diagonalized
leading to an energy spectrum consisting of two bands
\bea E_k^{\pm} = \pm 2 ~\sqrt{J_1^2 + J_2^2 + 2J_1 J_2 \cos k}. \label{en1}
\eea
Note that the band gap vanishes at $J_1= \pm J_2$ for $k=\pi$ and $0$
respectively, where the bands touch each other. It was shown in Ref.\
\onlinecite{feng} that this vanishing of the energy gap signals a topological
phase transition between the two phases of the model at $J_1 > J_2$ and
$J_1 < J_2$.

To study the quench of the system across this critical point, we
will now consider what happens when we evolve $J_1$ linearly in time
at a rate $1/\tau$ from $-\infty$ to $\infty$, keeping $J_2$ fixed at
some positive value: we take $J_1=J_2 t/\tau$. The ground states of
$H_{\rm 1D}$ in Eq.\ (\ref{kitham1d3}) have $\si_{2n}^x \si_{2n+1}^x =
1$ and $-1$ for $t=-\infty$ and $\infty$ respectively for all values
of $n$. In terms of the Hamiltonian in Eq. (\ref{kitham1d3}), the
ground and excited states for $J_1 \to - \infty$ are respectively given by
\bea \psi_{1k} &=& \frac{1}{\sqrt{2}} ~\left( \begin{array}{c} 1 \\ i
\end{array} \right) ~~{\rm and}~~ \psi_{2k} ~=~ \frac{1}{\sqrt{2}} ~\left(
\begin{array}{c} 1 \\ -i \end{array} \right). \label{wave} \eea
For $J_1 \to \infty$, the ground and excited states are given by
$\psi_{2k}$ and $\psi_{1k}$ respectively.

By a change of basis, one can rewrite Eq. (\ref{kitham1d3}) in the
form $H_{\rm 1D} = \sum_k \psi^{'\da}_k H'_k \psi'_k$ where
\bea H'_k ~=~ 2 ~\left( \begin{array}{cc} J_1 + J_2 \cos k & - J_2 \sin k \\
- J_2 \sin k & - J_1 - J_2 \cos k \end{array} \right). \label{kitham1d4} \eea
Note that unlike Eq.\ (\ref{kitham1d3}), the
off-diagonal elements of Eq.\ (\ref{kitham1d4}) do not change with
time if $J_2$ is held fixed. As a result, the problem of quench
dynamics is reduced to solving a standard Landau-Zener problem for
each momentum $k$ \cite{lz}. The density of defect formation $n$
can thus be found to be \cite{book1}
\bea n &=& \int_0^\pi ~\frac{dk}{\pi} ~p_k, \non \\
{\rm where} ~~p_k &=& e^{-2 \pi J_2 \tau \sin^2 k} \label{defect1} \eea
denotes the probability of the system to remain in the initial
($J_1 \to -\infty$) ground state for momentum $k$. For $J_2 \tau \gg 1$,
the contribution to $n$ comes mainly from the regions near $k=0$ and
$\pi$ where $p_k =1$. Thus one finds that in the slow quench regime
$n \simeq 1/\sqrt{J_2 \tau}$. Such a $1/\sqrt{\tau}$ scaling of
defect density conforms to the prediction of Ref.\
\onlinecite{anatoly1}. For the present case, it is easy to see from
Eq.\ (\ref{en1}), that the gap $\Delta(k)= E^+ (k) - E^- (k)$
vanishes linearly at the critical point both with the quench parameter
$J_1$ and with momentum around $k =0$ and $\pi$, so that $z\nu = z = 1$.
Thus, $n \sim 1/\tau^{d\nu/(z\nu +1)} = 1/\sqrt{\tau}$ in 1D.

\begin{figure}
\rotatebox{0}{\includegraphics*[width=\linewidth]{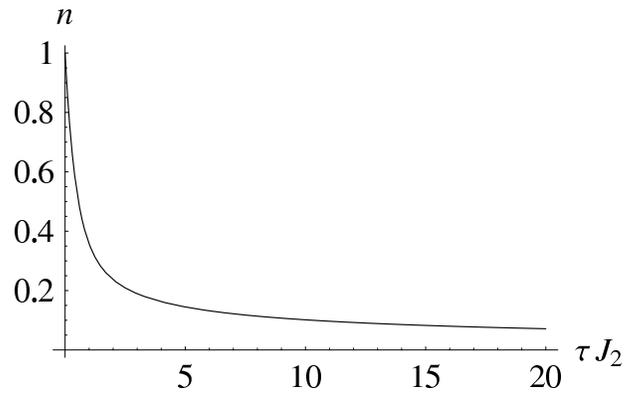}}
\caption{Defect density produced by quenching $J_1$ in $d=1$.}
\label{fig1} \end{figure}

A plot of the defect density as a function of the quench time $\tau$
is shown in Fig.\ \ref{fig1}. The plot confirms the expected result,
that the defect density is maximum for an infinite quench rate
($\tau \to 0$), when the system has no time to adjust to the
quench and remains in the old ground state leading to a normalized
defect density of $1$. As the rate of quench is decreased, $n$ decreases
quickly before settling down to a $1/\sqrt{\tau}$ behavior for large $\tau$.

It is useful to note that the Hamiltonian $H_k$ in Eq. (\ref{kitham1d3})
can also be written, after a suitable change of basis, as
\beq H'_k ~=~ 2 ~\left( \begin{array}{cc} J_- \sin (k/2) & - i J_+ \cos
(k/2)\\
i J_+ \cos (k/2) & - J_- \sin (k/2) \end{array} \right), \eeq
where $J_{\pm} = J_1 \pm J_2$. This form is useful if, for instance, one
wants to study the effect of quenching $J_-$ from $-\infty$ to $\infty$
keeping $J_+$ fixed.

\subsection{Correlation functions}
\label{1db}

Let us now consider how the system may be described at the final
time $t \to \infty$ when $J_1 = \infty$. In principle, the time
evolution of the system is unitary, so that it will always be a pure
state. However, for each momentum $k$, the wave function is given by
$\sqrt{1 - p_k} \psi_{2k} e^{-iE_k^2 t} ~+~ \sqrt{p_k} \psi_{1k}
e^{-iE_k^1 t}$, where $E_k^{1,2} = \pm \infty$. As a result, the
final density matrix of the system will have off-diagonal terms
involving $\psi_{2k}^* \psi_{1k}$ and $\psi_{1k}^* \psi_{2k}$ which
vary extremely rapidly with time; their effects on physical
quantities will therefore average to zero. Hence the final density
matrix is effectively diagonal like that of a mixed state
\cite{levitov}, where the diagonal entries are time-independent as
$t \to \infty$ and are given by $1 - p_k$ and $p_k$. Such a density
matrix is associated with an entropy which we will discuss in Sec.
\ref{ent} in the context of 2D Kitaev model.

Using the above density matrix, we will now compute the correlation
functions corresponding to the operators $O_r = i b_n a_{n+r}$, where $r$
is an integer. In terms of the spins, as can be seen from Eq.\ (\ref{maj1d}),
the operator $O_r$ can be written as
\bea O_0 &=& \si_{2n}^x \si_{2n+1}^x, ~~~~ O_1 ~=~ \si_{2n+1}^y \si_{2n+2}^y,
\non \\
O_r &=& \si_{2n+1}^y ~\left( \prod_{j=2n+2}^{2n+2r-1} \si_j^z \right)~
\si_{2n+2r}^y ~~{\rm for} ~~ r \ge 2, \non \\
&=& \si_{2n+2r}^y ~\left( \prod_{j=2n+2r+1}^{2n} \si_j^z \right)~
\si_{2n+1}^y ~~ {\rm for} ~~r \le -1. \non \\
& & \eea
We will calculate the expectation values of these operators shortly. In
principle, one can also consider expectation values of the operators $i a_n
a_{n+r}$ and $i b_n b_{n+r}$; however a direct calculation shows that
these vanish if $r \ne 0$.
Further, for the Kitaev model, it has been shown that the spin-spin
correlations between sites lying on different bonds vanish, {\it i.e.}, $\la
\si_{2n}^x \si_{2n+r}^x \ra =0$ for $|r| \ge 2$ \cite{baskaran}. Therefore
$\la O_r \ra$ are the only non-vanishing spin-correlators of the model
\cite{nussinov1}.

To compute $\la O_r \ra$ we note that $O_r$ can be expressed in
terms of the fermion operators $a_k$ and $b_k$. This will in general
involve summations over two different momenta $k$ and $k'$. However,
when $\la O_r \ra$ is computed in a direct product of states
involving $a_k$ and $b_k$, only terms in which $k' = k$ will
contribute. In the limit $N \to \infty$, the relevant part of $O_r$
which contributes to the correlation function can be written as \beq
O_r ~=~ - ~\frac{4i}{N} ~\sum_{k=0}^\pi ~[~ b_k^\da a_k e^{ikr} ~-~
a_k^\da b_k e^{ikr} ~]. \label{corr2} \eeq

\begin{figure}
\rotatebox{0}{\includegraphics*[width=\linewidth]{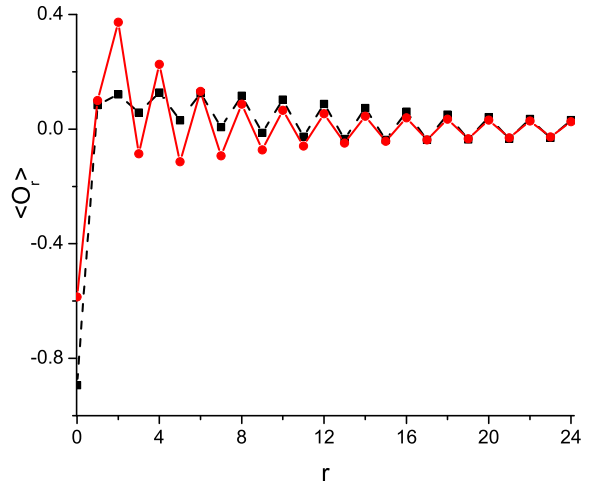}}
\caption{Plot of correlation function $\la O_r \ra$ as a function of
$r$ for $J_2 \tau = 10$ (red circles and red solid line) and $J_2
\tau =1$ (black squares and black dashed line). $\la O_r \ra$ shows
a damped oscillatory behavior as a function of $r$.} \label{fig2} \end{figure}

Using the wave functions given in Eq. (\ref{wave}), we find that
\beq \la O_r \ra ~=~ \pm ~\int_0^\pi dk \cos (kr) ~=~ \pm \de_{r,0}, \eeq
where the $+$ and $-$ signs refer to the ground states of $J_1 = - \infty$
and $\infty$ respectively. This is expected since $\si_{2n}^x \si_{2n+1}^x =
\pm 1$ while all other correlations vanish in those two states.
Finally, after quench, in a state in which we have a mixture of the ground
and excited states of $J_1 = \infty$ with probabilities $1 - p_k$ and $p_k$
respectively, we find that
\beq \la O_r \ra ~=~ - ~\de_{r,0} ~+~ \frac{2}{\pi} ~\int_0^\pi dk ~p_k ~\cos
(kr). \label{int1} \eeq
A plot of $\la O_r \ra$ as a function of $r$ is shown for representative
values of $J_2 \tau=1,10$ in Fig.\ \ref{fig2}. We find that $\la O_r \ra$
shows a damped oscillatory behavior.
Note that since $\la O_r \ra = -\delta_{r,0}$ for the ground state
of $H_{\rm 1D}$ for $J_1 \to \infty$, the plot of $\la O_r \ra$ in
the state of the system after the quench provides a direct
measurement of the spatial extent of the correlation between the
defects generated during the quench.

For $J_2 \tau \gg 1$, the dominant contribution in the integral in
Eq. (\ref{int1}) comes from the regions near $k = 0$ and $\pi$ as can be seen
from the expression for $p_k$ in Eq.\ (\ref{defect1}). One can combine these
two regions, and write the expression in (\ref{int1}) approximately as
\bea \la O_r \ra &=& - ~\de_{r,0} ~+~ \frac{2}{\pi} ~\int_0^\infty
dk ~e^{-2 \pi J_2 \tau k^2} \non \\
& & ~~~~~~~~~~~~~~~~~~~~\times ~[\cos (kr) ~+~ \cos \{(\pi - k) r \}] \non \\
&=& - ~\de_{r,0} ~+~ \frac{1 ~+~ (-1)^r}{\pi} ~\frac{e^{-r^2 / (8 \pi J_2
\tau)}}{\sqrt{2 J_2 \tau}}. \label{or} \non \\ \eea
Note that this vanishes if $r$ is odd. For a given value of $J_2 \tau$,
the expression in Eq. (\ref{or}) decreases with increasing $r$,
particularly for $r > \sqrt{8 \pi J_2 \tau}$. On the other hand, for
a given large value of $r$, Eq. (\ref{or}) has a maximum at $\tau =
r^2 /(4 \pi J_2)$. The fact that the crossover in both cases occurs
around $r \sim \sqrt{4 \pi J_2 \tau}$ signals the fact that the
associated length scale for the defect correlation function is of
order $\sqrt{4 \pi J_2 \tau}$.

\subsection{Sum rule}
\label{sumrule}

There is a sum rule that we can write down for $\left<O_r\right>$.
{}From Eq. (\ref{int1}), we see that \beq O_{total} ~\equiv~
\sum_{r=-\infty}^\infty ~\la O_r \ra ~=~ -1 ~+~ 2 p_0, \label{sum}
\eeq where we have used the identity $\sum_r e^{ikr} = 2\pi \de (k)$
for $-\pi < k < \pi$. Going back to Eq. (\ref{kitham1d4}), we see
that for $k=0$, the Hamiltonians at different times commute with
each other irrespective of how $J_1$ is varied in time from
$-\infty$ to $\infty$. This means that if we start with the ground
state of $J_1 = - \infty$, no transition will occur at any time, and
we will have $p_0 = 1$. Eq. (\ref{sum}) then implies that $O_{total} = 1$.

\section{2D Kitaev model}
\label{2d}

\subsection{Defect density}
\label{2da}

When $J_3 \ne 0$, the Kitaev model with Hamiltonian given by Eq.\
(\ref{kham1}) describes a spin model on a hexagonal 2D lattice.
Usually spin models are not exactly solvable in two dimensions. One
of the main properties of the Kitaev model which makes it
theoretically attractive is that, even in 2D, it can be mapped onto
a non-interacting fermionic model by a suitable Jordan-Wigner
transformation \cite{kitaev1,feng,nussinov1,nussinov2}. In terms of
the Majorana fermions $a_{jl}$ and $b_{jl}$ one can write
\bea a_{jl} &=& \left(
\prod_{i=-\infty}^{j-1} ~\si_{il}^z \right) ~
\si_{jl}^y ~~{\rm for}~{\rm ~ even }~ j+l, \non \\
b_{jl} &=& \left( \prod_{i=-\infty}^{j-1} ~\si_{il}^z \right)
~\si_{jl}^x ~~ {\rm for}~{\rm ~odd }~ j+l. \label{maj2d} \eea
Such a transformation maps the spin Hamiltonian $H$ in Eq.\ (\ref{kham1})
to a fermionic Hamiltonian given by \bea H_{\rm 2D} &=& i
~\sum_{\vn} ~[J_1 ~b_{\vn} a_{{\vn} - {\vec M}_1} ~+~
J_2 ~ b_{\vn} a_{{\vn} + {\vec M}_2} \non \\
&& ~~~~~~~~~+~ J_3 D_{\vn} ~b_{\vn} a_{\vn}], \label{kitham2d1} \eea
where $\vn = {\sqrt 3} {\hat i} ~n_1 + (\frac{\sqrt 3}{2} {\hat i} +
\frac{3}{2} {\hat j} ) ~n_2$ denote the midpoints of the vertical
bonds. Here $n_1, n_2$ run over all integers so that
the vectors $\vn$ form a triangular lattice whose vertices
lie at the centers of the vertical bonds of the underlying honeycomb
lattice; the Majorana fermions $a_{\vn}$ and $b_{\vn}$ sit at the
top and bottom sites respectively of the bond labeled $\vn$. The
vectors ${\vec M}_1 = \frac{\sqrt 3}{2} {\hat i} +
\frac{3}{2} {\hat j}$ and ${\vec M}_2 = \frac{\sqrt 3}{2} {\hat i} -
\frac{3}{2} {\hat j}$ are spanning vectors for the reciprocal lattice, and
$D_{\vn}$ can take the values $\pm 1$ independently for each $\vn$. The
crucial point that makes the solution of Kitaev model feasible is that
$D_{\vn}$ commutes with $H_{\rm 2D}$, so that all the eigenstates of
$H_{\rm 2D}$ can be labeled by specific values of $D_{\vn}$. It has been
shown that for any value of the parameters $J_i$, the ground state of the
model always corresponds to $D_{\vn}=1$ on all the bonds. Since $D_{\vn}$ is
a constant of motion, the dynamics of the model starting from any ground
state never takes the system outside the manifold of states with $D_{\vn}=1$.

For $D_{\vn}=1$, it is straightforward to diagonalize $H_{\rm 2D}$ in momentum
space. We define Fourier transforms of the Majorana operators $a_{\vn}$ as
\beq a_{\vn} ~=~ \sqrt{\frac{4}{N}} ~\sum_{\vk} ~[~ a_{\vk} ~e^{i\vk \cdot
\vn} ~+~ a_{\vk}^\da ~ e^{-i\vk \cdot \vn} ~] \label{fourier2} \eeq
(and similarly for $b_{\vn}$), where $N$ is the number of sites (assumed to be
even, so that the number of unit cells $N/2$ is an integer), and the sum over
$\vk$ extends over half the Brillouin zone of the 2D hexagonal lattice. We then
obtain $H_{\rm 2D} = \sum_{\vk} \psi_{\vk}^\da H_{\vk} \psi_{\vk}$, where
$\psi_{\vk}^\da =(a_{\vk}^\da , b_{\vk}^\da)$, and
$H_{\vk}$ can be expressed in terms of Pauli matrices $\si^{1,2,3}$ as
\bea H_{\vk} &=& 2 ~[J_1 \sin ({\vk} \cdot {\vec M}_1) ~-~ J_2
\sin ({\vk} \cdot {\vec M}_2)] ~\si^1 \non \\
& & +~ 2 ~[J_3 ~+~ J_1 \cos ({\vk} \cdot {\vec M}_1) ~+~ J_2 \cos
({\vk} \cdot {\vec M}_2)] ~\si^2 . \label{ham2} \non \\ \eea
The energy spectrum of $H_{\rm 2D}$ therefore consists of two bands
with energies
\bea E_{\vk}^\pm &=& \pm ~2 ~[(J_1 \sin ({\vk} \cdot {\vec
M}_1) - J_2 \sin ({\vk} \cdot {\vec M}_2))^2 \non \\
&& ~~~~+ (J_3 + J_1 \cos ({\vk} \cdot {\vec M}_1) + J_2 \cos ({\vk}
\cdot {\vec M}_2))^2 ]^{1/2} . \non \\ \label{hk1} \eea
We note for $|J_1-J_2|\le J_3 \le (J_1+J_2)$, these bands touch each
other so that the energy gap $\Delta_{\vk} =
E_{\vk}^+ - E_{\vk}^-$ vanishes for special values of $\vk$
leading to the gapless phase of the model \cite{kitaev1,feng,lee,nussinov1}.

We will now quench $J_3(t) =J t/\tau$ at a fixed rate $1/\tau$,
from $-\infty$ to $\infty$, keeping $J$, $J_1$ and $J_2$ fixed at some
non-zero values; we have introduced the quantity $J$ to fix the
scale of energy. We note that the ground states of $H_{\rm 2D}$
corresponding to $J_3 \to -\infty(\infty)$ are gapped and
have $\si_{j,l}^z \si_{j,l+1}^z = 1(-1)$ for all lattice sites
$(j,l)$. To study the state of the system after the quench, we first
note that after an unitary transformation $U= \exp(-i \si_1
\pi/4)$, one can write $H_{\rm 2D} = \sum_{\vk} \psi_{\vk}^{'\da}
H'_{\vk} \psi'_{\vk}$, where $H'_{\vk} = UH_{\vk} U^\da$ is given by
\bea H'_{\vk} &=& 2 ~[J_1 \sin ({\vk} \cdot {\vec M}_1) ~-~ J_2
\sin ({\vk} \cdot {\vec M}_2)] ~\si^1 \non \\
& & + ~2 ~[J_3(t) +J_1 \cos ({\vk} \cdot {\vec M}_1) + J_2 \cos
({\vk} \cdot {\vec M}_2)] ~\si^3 . \non \\ \eea
Hence the off-diagonal elements of $H'_{\vk}$ remain time
independent and the problem of quench dynamics reduces to a
Landau-Zener problem for each ${\vk}$. The defect density can then
be computed following a standard prescription \cite{lz}
\bea n &=& \frac{1}{A} ~\int_{\vk} ~d^2 \vk ~p_{\vk}, \non \\
p_{\vk} &=& e^{ - 2 \pi \tau ~[J_1 \sin ({\vk} \cdot {\vec M}_1)- J_2
\sin ({\vk} \cdot {\vec M}_2)]^2/J}, \label{defect2d} \eea
where $A = 4\pi^2 /(3\sqrt{3})$ denotes the area of half the Brillouin zone
over which the integration is carried out. Since the integrand in Eq.
(\ref{defect2d}) is an even function of $\vk$, one can extend the region of
integration over the full Brillouin zone. This region can be chosen
to be a rhombus with vertices lying at $(k_x,k_y)= (\pm 2\pi /
\sqrt{3}, 0)$ and $(0,\pm 2\pi /3)$. Introducing two independent integration
variable $v_1, v_2$, each with a range $0\le v_1,v_2 \le 1$, one finds that
\bea k_x &=& 2\pi ~\frac{v_1 + v_2 -1}{\sqrt 3}, \quad k_y = 2\pi ~
\frac{v_2 - v_1}{3}. \eea
Such a substitution covers the rhombus uniformly and facilitates the
numerical integration necessary for computing $n$.

\begin{figure}
\rotatebox{0}{\includegraphics*[width=\linewidth]{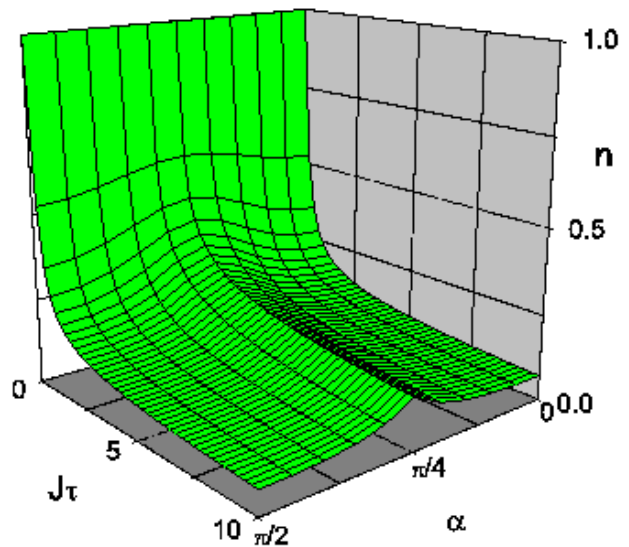}}
\caption{Plot of defect density $n$ as a function of the quench time
$J \tau$ and $\alpha = \tan^{-1} (J_2/J_1)$. The density of defects is
maximum at $J_1=J_2$.} \label{fig3} \end{figure}

A plot of $n$ as a function of the quench time $J \tau$ and $\alpha =
\tan^{-1} (J_2/J_1)$ (we have taken $J_{1[2]}= J \cos(\alpha)[\sin(\alpha)]$)
is shown in Fig.\ \ref{fig3}. We note
that the density of defects produced is maximum when $J_1=J_2$. This
is due to the fact that the length of the gapless line through which
the system passes during the quench is maximum at this point. This
allows the system to remain in the non-adiabatic state for the maximum time
during the quench, leading to the maximum density of defects. For $J_1/J_3
>2J_2/J_3$, the system does not pass through a gapless phase during the
quench, and the defect production is exponentially suppressed.

For sufficiently slow quench $2 \pi J \tau \gg 1$, $p_{\vk}$ is
exponentially small for all values of ${\vk}$ except in the region
near the line \beq J_1 ~\sin ({\vk} \cdot {\vec M}_1) ~-~ J_2 \sin
({\vk} \cdot {\vec M}_2) ~=~ 0, \label{line} \eeq and the
contribution to the momentum integral in (\ref{defect2d}) comes from
values of $\vk$ close to this line of zeroes. We note that the line
of zeroes where $p_{\vk}=1$ precisely corresponds to the zeroes of
the energy gap $\Delta_{\vk}$ as $J_3$ is varied for a fixed $J_2$
and $J_1$. Thus the system becomes non-adiabatic when it passes
through the intermediate gapless phase in the interval $|J_1-J_2|\le
J_3(t) \le (J_1+J_2)$. It is then easy to see, by expanding
$p_{\vk}$ about this line that in the limit of slow quench, the
defect density scales as $n \sim 1/\sqrt{\tau}$. We note that the
scaling of the defect density with the quench rate in a quench where
the system passes through a critical {\it line} in momentum space is
different from the situation where the quench takes the system
through a critical {\it point}. In the latter case, for the Kitaev
model which has $z=\nu=1$, Ref.\ \onlinecite{anatoly1} predicts a
defect density $n \sim 1/\tau$ for $d=2$. Thus the defect density
crucially depends on the dimensionality of the critical surface
through which the system passes during the quench. This observation
leads to a simple but general conclusion which we present below.

Consider a $d$-dimensional model with $z=\nu=1$ described by a Hamiltonian
\bea H_{d} &=& \sum_{\vk} \psi^\da_{\vk} \left( \begin{array}{cc}
\epsilon(\vk,t) & \Delta(\vk) \\
\Delta^{\ast}(\vk)& - \epsilon(\vk,t) \end{array} \right)
\psi_{\vk}, \label{hd} \eea where
$\epsilon(\vk,t)=\epsilon(\vk)t/\tau$. Now let us assume that the
quench takes such a system through a critical surface of $d-m$
dimensions. The defect density for a sufficiently slow quench can be
expressed as \cite{lz,book1} \bea n &=& \frac{1}{A} ~\int_{\rm BZ}
d^d k ~p(\vk), ~~{\rm where} ~~p(\vk)
=e^{-\pi \tau f(\vk)}, \non \\
&\simeq & \frac{1}{A} ~\int_{\rm BZ} d^d k ~ \exp [~-\tau \sum_{\alpha
\beta=1,m} g_{\alpha \beta} k_{\alpha} k_{\beta}] \non \\
& \sim & 1/\tau^{m/2} , \label{genres} \eea
where $p_{\vk}$ is the defect probability for momentum $\vk$,
$f (\vk)=|\Delta(\vk)|^2/|\epsilon(\vk)|$ vanishes on the $d-m$
dimensional critical surface, $\alpha, \beta$ denote one of the $m$
orthogonal directions to the critical surface and $g_{\alpha \beta}
= (\partial^2 f(\vk)/\partial k_{\alpha}
\partial k_{\beta})_{\vk \in {\rm critical ~surface}}$. We note that
this result depends only on the property that $f(\vk)$ has to vanish
on a $d-m$ dimensional surface, and not on the precise nature of
$f(\vk)$. For $m=d$, where the quench takes the system through a critical
point, our results coincide with that of Ref.\ \onlinecite{anatoly1}.

Finally we generalize our arguments for models where the $d-m$
dimensional hypersurface is characterized by correlation length
exponent $\nu$ and dynamical critical exponent $z$. Let us assume
that the system is described by a Hamiltonian $H[\lambda(t)]$ with
quasi-energy eigenvalues $E(\vk,t)$ and that the time evolution of
the parameter $\lambda(t)=\lambda_0 (t/\tau)$ takes the system
through the critical point $\lambda_0=\lambda_c$ at $t=t_0$. First
we note that for large $\tau$, a non-vanishing probability of defect
formation requires the non-adiabaticity condition $|\Delta(\vk)|^2
\sim |\partial E(\vk,t)/\partial t|$ [\onlinecite{anatoly1}]. Also,
since $\partial E(\vk,t)/\partial t =(\partial E(\vk,t)/\partial
\lambda)\tau^{-1}$ and near the critical point $E \sim \lambda^{z\nu}$, we get
\bea \Delta^2 \sim \tau^{-1} \lambda^{z\nu-1} \label{drule1} \eea
Further, as shown in Ref.\ \onlinecite{anatoly1}, near any point on
the critical surface, quite generally, one has $\Delta \sim |k|^z$,
$\lambda \sim k^{1/\nu}$ and $ k \sim 1/\tau^{\nu/(z\nu +1)}$.
Using these relations we find from Eq.\ \ref{drule1} that on any
point near the gapless surface
\bea \Delta \sim 1/\tau^{z\nu/(z\nu+1)} \label{drule2} \eea
Next, let us consider the available phase space for formation of
defects. When the quench takes the system through a $d-m$
dimensional hypersurface in momentum space, the available phase
space is $\Omega \sim k^m \sim \Delta^{m/z}$. Since this available
phase space is directly proportional to the defect density
\cite{anatoly1}, we find, using Eq.\ (\ref{drule2}),
\bea n \sim \Omega \sim \Delta^{m/z} \sim 1/\tau^{m \nu/(z\nu+1)}
\label{gquench1} \eea
This generalizes the scaling law for defect density to arbitrary critical
systems. Note that for $z=\nu=1$, we recover our earlier result $n \sim
1/\tau^{m/2}$ (Eq.\ (\ref{genres})). For $m=d$, which represents quench
through a critical point, we also recover the result of Ref.\
\onlinecite{anatoly1} ($ n \sim 1/\tau^{d\nu/(z\nu+1)}$) as a special case.

\subsection{Defect Correlation}
\label{2db}

The calculation of the correlation function can be accomplished
along similar lines as in 1D. First, we define the operators \beq
O_{\vcr}^{\rm 2D} ~=~ i b_{\vn} a_{\vn + \vcr}. \label{kitop1} \eeq
In terms of the spin operators, we have $O_{\vec 0}^{\rm 2D} =
\si_{j,l}^z \si_{j,l+1}^z$. For $\vcr \ne {\vec 0}$, $O_{\vcr}^{\rm
2D}$ can be written as a product of spin operators going from a $b$
site at $\vn=(j,l)$ to an $a$ site at $\vn + \vcr = (j',l')$: the
product will begin with a $\si^x$ or $\si^y$ and end with a $\si^x$
or $\si^y$ with a string of $\si^z$'s in between, where the choice
of the initial and final $\si$ matrices depends on whether the
values of $j+l$ and $j'+l'$ are even or odd.
Note that $O_{\vcr}^{\rm 2D}$ for $\vcr \ne {\vec 0}$ measures
correlation between the defects produced during the quench. In
particular, a plot of the correlation function $\la O_{\vcr}^{\rm
2D} \ra$ versus $\vcr$ in the defect ground state provides an
estimate of the shape and spatial extent of the defect correlations
produced during the quench. Note that $(O_{\vcr}^{\rm 2D})^2 = 1$,
so that all the moments of $O_{\vcr}^{\rm 2D}$ can be found
trivially: $\la (O_{\vcr}^{\rm 2D})^n \ra = \la O_{\vcr}^{\rm 2D}
\ra$ if $n$ is odd and $=1$ if $n$ is even.

$O_{\vcr}^{\rm 2D}$ can be written in terms of the Majorana
fermion operators $a_{\vk}$ and $b_{\vk}$; this again involves a sum
over two different momenta $\vk$ and $\vk'$. However, the
expectation value of $O_{\vcr}^{2D}$ in a direct product of states
involving $\vk$ only gets a contribution from terms in which $\vk' = \vk$.
It turns out that the relevant part of $O_{\vcr}^{2D}$ contributes
to the expectation values can be written as,
\beq O_{\vcr}^{2D} ~=~ \frac{4i}{N} ~\sum_{\vk} ~[ b_{\vk}^\da a_{\vk}
e^{i\vk \cdot \vcr} ~-~ a_{\vk}^\da b_{\vk} e^{-i\vk \cdot \vcr}]. \eeq

The ground state and excited states for $J_3 = - \infty$ are given by
$\psi_{1\vk}$ and $\psi_{2\vk}$ respectively, while the two states are
interchanged for $J_3 = \infty$. Using Eq.\ (\ref{wave}), we find that
\beq \la O_{\vcr}^{\rm 2D} \ra ~=~ \pm \frac{4}{N} ~\sum_{\vk} ~\cos (\vk
\cdot \vcr), \eeq
where the $+$ and $-$ signs refer to the ground states
of $J_3 = - \infty$ and $\infty$ respectively. This confirms our
earlier expectation that in the ground states of $J_3 \to -
\infty (\infty)$, $\la O_{\vcr}^{\rm 2D} \ra = \pm \delta_{\vcr, {\vec 0}}$.
Finally, in the state after quench, in which we have a mixture
of the ground and excited states of $J_3 = \infty$ with
probabilities $1 - p_{\vk}$ and $p_{\vk}$ respectively, we find that
\bea \la O_{\vcr}^{\rm 2D} \ra &=& - ~\de_{\vcr,{\vec 0}} ~+~ \frac{2}{A} ~
\int ~d^2 \vk ~ p_{\vk} ~\cos (\vk \cdot \vcr), \label{int2} \eea
where the integral over momentum runs over half the Brillouin zone with
area $A$. Note that the full Brillouin zone as well as $p_{\vk}$ remains
invariant under a reflection through the point $\vk = (\pi/\sqrt{3}, 0)$:
$k_x \to 2\pi/\sqrt{3} - k_x$, $k_y \to - k_y$. However, $\cos (\vk
\cdot {\vcr})$ changes by a factor of $(-1)^{n_2}$, if the components of
$\vcr$ are given by $x=\sqrt{3}(n_1 +n_2/2)$ and $y=3n_2/2$. Hence,
$\la O_{\vec{r}}^{\rm 2D} \ra =0$ for odd values of $n_2$.

\begin{figure}
\rotatebox{0}{\includegraphics*[width=\linewidth]{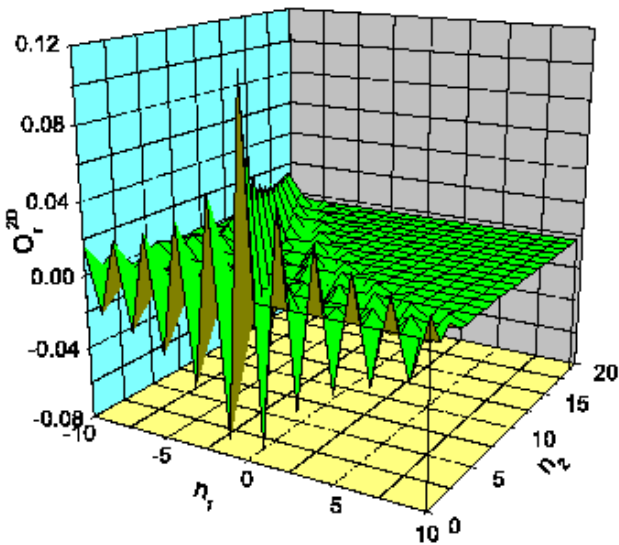}}
\caption{Plot of $O_{\vcr}^{\rm 2D}$ sans the delta function peak at the
origin for $J_1=J_2=J$ and $J \tau=10$ as a function of $n_1$ and $n_2$
(see text for details). The spatial anisotropy of the defect correlation
function is clearly evident even for $J_1=J_2$.} \label{fig4} \end{figure}

For large values of $\tau$, substituting the expression in Eq.\
(\ref{defect2d}) in the above integral, we find that the dominant
contribution comes from the region near the line given in Eq.
(\ref{line}). Thus at every point $\vk_0$ lying on that line, we can
introduce variables $k_\parallel$ and $k_\perp$ which vary along the
line and perpendicular to it along the directions ${\hat
n}_\parallel$ and ${\hat n}_\perp$ respectively. Close to $\vk_0$,
the integrand in Eq.\ (\ref{int2}) will take the form $\exp [-a\tau
k_\perp^2 \pm i (\vk_0 + k_\parallel {\hat n}_\parallel + k_\perp
{\hat n}_\perp) \cdot \vcr]$, where $a$ is a number of order 1 whose
value depends on $\vk_0$. The integral over $k_\perp$ will give a
factor of $\exp \left[-(\vcr \cdot {\hat n}_\perp)^2
/(4a\tau)\right]/\sqrt{a \tau}$. Thus we find that the density of
defects is of order $1/\sqrt{\tau}$ in accordance with Eq.\
(\ref{genres}). This also leads us to expect that the spatial range
of the defect correlation should go as $\sqrt{\tau}$.

Next we consider the shape of the defect correlation function. For
this purpose, we evaluate Eq.\ (\ref{int2}) numerically so as to
obtain the $\vcr$ dependence of $\la O_{\vcr}^{\rm 2D} \ra$. In
general, we expect the correlation will be anisotropic in space if
$J_1/J_2 \gg 1$ or $\ll 1$ which can be most easily seen from the
fact that setting $J_1=0$ or $J_2=0$ leads to the 1D result derived
in Sec.\ \ref{1db}. A plot of the correlation function $\la
O_{\vcr}^{\rm 2D} \ra$, without the delta function peak at
${\vcr}=0$, and as a function of $n_1$ and $n_2$, where
$x=\sqrt{3}(n_1 +n_2/2)$ and $y=3n_2/2$ is shown in Fig.\
\ref{fig4}. In this plot, we have omitted the delta function
contribution to $\la O_{\vcr=0}^{\rm 2D}\ra$ in order to make the
correlations at $\vcr \ne {\vec 0}$ visible. In the $x$ direction,
the correlations oscillate; the amplitude of oscillations decays
monotonically with $x$, in a qualitatively similar manner to the 1D
correlation function $O_r$ shown in Fig.\ \ref{fig1} for $y=n_2=0$.
The correlations decay in a monotonic way with $y$ for $x = n_1 +
n_2/2 = 0$ (along the straight line at an angle $\ta =
\tan^{-1}(-0.5)$ in the figure). Thus the correlations behave quite
anisotropically even for $J_1=J_2$.

\begin{figure}
\rotatebox{0}{\includegraphics*[width=\linewidth]{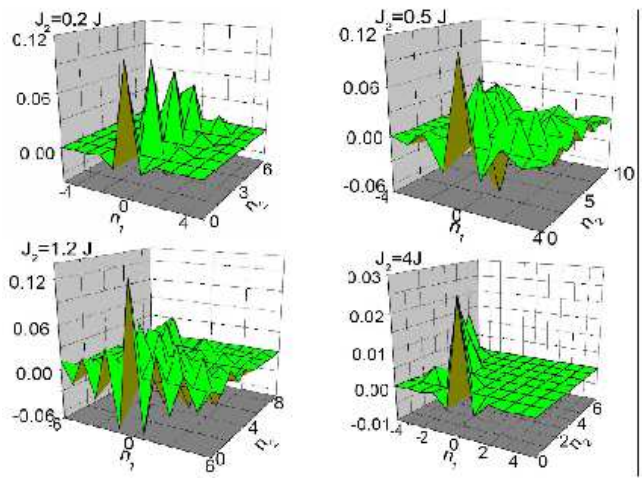}}
\caption{Plot of $\left<O_{r}^{\rm 2d}\right>$ sans the delta
function peak at the origin as a function of $\vcr$ for several
representative values of $J_2/J$ for $J_1=J$ and $J\tau=5$. The plot
displays the change in the shape of defect correlation function as a
function of $J_2/J_1$ (see text for details).} \label{fig5} \end{figure}

We now aim at obtaining an understanding of the variation of the
spatial dependence of $\left<O_r^{\rm 2d}\right>$ with the
parameters $J_1$ and $J_2$. Such a variation can be analytically
understood by noting that for $J\tau \gg 1$, the maximum
contribution to $\left< O_{\vcr} \right>$ comes from around the
wave vector ${\vec k}_0$ for which $p({\vec k}_0)=1$. For $J_2 \gg
(\ll) 1$, this occurs when $\sin[{\vec k}\cdot {\vec M}_2({\vec
M}_1)] =0$ which yields ${\vec k}_0 = \pi (\sqrt{3} {\hat i} \mp
{\hat j})/2$. The maximum contribution to $\left<O_{\vcr}^{\rm 2D}
\right>$ occurs where $\cos({\vec k}_0 \cdot {\vec n})$ is
maximum, ${\it ie.}$, ${\vec k}_0 \cdot {\vec n}=0$. Thus for $J_2
\gg (\ll) J_1$, $\left<O_{\vcr}^{\rm 2D}\right>$ is expected to be
maximal along the lines $n_1 +n_2=0 (n_2=0)$ in the $n_1-n_2$ plane.
This expectation is confirmed as seen in Fig.\ \ref{fig5} which
shows $\left<O_r^{\rm 2d}\right>$ for several representative values
of $J_2/J$ for a fixed $J_1=J$ and $J\tau=5$. We find that
$\left< O_{\vcr} \right>$ is maximal along $n_2=0 (n_1+n_2=0)$ line
for $J_2=5(0.25)J_1$. This clearly shows that the defects produced
in the quench will be highly anisotropic in this limits. For
intermediate values of $J_1$ and $J_2$, the anisotropy in
$\left<O_r^{\rm 2d}\right>$ can be similarly deduced by first
finding ${\vec k}_0$ for which $p_{\vec k_0}=0$ and then computing
${\vec n}$ for which ${\vec k}_0 \cdot {\vec n}$ vanishes. The
gradual evolution of the shape of $\left<O_r^{\rm 2d}\right>$ as we
go from the limit $J_2 \ll J_1$ to the limit $J_2 \ll J_1$ can be
seen from in Fig.\ \ref{fig5}.

\begin{figure}
\rotatebox{0}{\includegraphics*[width=\linewidth]{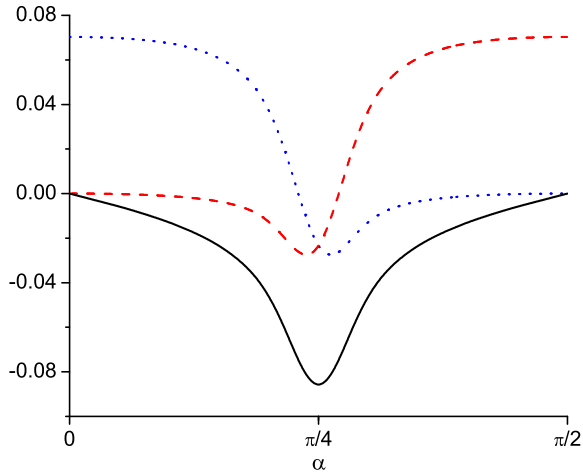}}
\caption{Plot of $\left<O_{r}^{\rm 2d}\right>$ (sans the delta
function peak) at representative points $(-1,0)$ on the $x$ axis
(black solid line) $(0,2)$ on the $y$ axis (blue dotted line) and
$(2,-2)$ along $-45^{\circ}$ in the $n_1-n_2$ plane (red dashed
line) as a function of $\alpha = \tan^{-1} (J_2/J_1)$ for fixed
$J^2=1$.} \label{fig7} \end{figure}

To obtain a more detailed picture of the spatial anisotropy of the
defect correlations as a function of $J_1/J_2$ we define a parameter
$\alpha$: $J_{1[2]} = J \cos(\alpha)[\sin(\alpha)]$. A variation of
$\alpha$ therefore changes the ratio $J_1/J_2$ from $0$ to $\infty$
while fixing $J_1^2+J_2^2=J^2=1$. The plot of $\left<O_{\vcr}^{\rm 2d}
\right>$ at points $(n_1,n_2)=(-1,0)$ (on the $x$ axis of the
$n_1-n_2$ plane), $(n_1,n_2)=(2,-2)$ (along the $-45^{\circ}$ line
in the $n_1-n_2$ plane) and $(n_1,n_2)=(0,-2)$ (on the $y$ axis of
the $n_1-n_2$ plane) as a function of $\alpha$ shown in Fig.\
\ref{fig7} clearly reveal the nature of the anisotropy of the
correlation function. We find that as the ratio of $J_1/J_2 =
\tan(\alpha)$ is varied from $0$ to $\infty$, the correlation on the
representative point $(1,0)$ along the $x$ axis increases till it
reaches the point $J_1=J_2$ ($\alpha=\pi/4$) and then decays to $0$
as $\alpha$ approaches $\pi/2$. This signifies that the correlation
along the $x$ axis in the $n_1-n_2$ plane becomes maximum for
$J_1=J_2$. On the other hand, for the representative point $(0,2)$
on the $y$ axis and $2,-2$ along the line with slope $-45^{\circ}$,
the correlation becomes maximum when $J_2 \ll J_1$ ($\alpha=0$) and
$J_2 \gg J_1$ ($\alpha=\pi/2$) respectively, as expected from Fig.\
\ref{fig5}. This lead us to conclude that the spatial anisotropy of
the defect correlation function $\left<O_{\vcr}^{\rm 2d}\right>$
depends crucially on the ratio of $J_1/J_2$.

Finally we note that we can obtain a measure of the spatial extent
of the defect correlation function by calculating
\beq \la \vcr^2 \ra ~\equiv~ \sum_{\vcr} ~\vcr^2 ~\la O_{\vcr}^{\rm 2D} \ra .
\eeq
To evaluate this, we first rewrite Eq. (\ref{int2}) as
\bea \la O_{\vcr}^{\rm 2D} \ra &=& - ~\de_{\vcr,{\vec 0}} ~+~ \frac{1}{A} ~
\int ~d^2 \vk ~ p_{\vk} ~e^{i\vk \cdot \vcr}, \label{int3} \eea
where the integral now runs over the entire Brillouin zone.
We then note that $\vcr^2 e^{i\vk \cdot \vcr} = - \nabla^2_{\vk} e^{i\vk
\cdot \vcr}$, integrate by parts in Eq. (\ref{int3}) so as to make
$\nabla^2_{\vk}$ act on $p_{\vk}$, and use the identity $\sum_{\vcr}
e^{i\vk \cdot \vcr} = 2 A \de^2 (\vk)$, to obtain $\la \vcr^2 \ra =
-2 (\nabla^2_{\vk} p_{\vk})_{\vk = {\vec 0}} = 24 \pi \tau (J_1^2 +
J_2^2 + J_1 J_2)/J$. This shows that the spatial extent of $\la
O_{\vcr}^{\rm 2D} \ra $ grows as $\sqrt \tau$ for large $\tau$. [Eq.
(\ref{or}) shows that we get the same behavior in 1D.] Finally, we
can get an idea of the spatial anisotropy of $\la O_{\vcr}^{\rm 2D}
\ra$ by computing \beq \la \vcr^2 \ra_\ta ~\equiv~ \sum_{\vcr} ~(x
\cos \ta + y \sin \ta)^2 ~ \la O_{\vcr}^{\rm 2D} \ra , \eeq where
$\vcr = (x,y)$, and $\ta$ denotes a direction along which the
spatial extent is being calculated. By writing $(x \cos \ta + y \sin
\ta)^2 e^{i\vk \cdot \vcr} = - (\cos \ta \partial /\partial k_x +
\sin \ta \partial /\partial k_x)^2 e^{i\vk \cdot \vcr}$, we can
prove that $\la \vcr^2 \ra_\ta = 6 \pi \tau [(J_1 - J_2) \cos \ta +
\sqrt{3} (J_1 + J_2) \sin \ta]^2 /J$. We see that $\la \vcr^2
\ra_\ta$ has a marked dependence on $\ta$; in fact, it vanishes in
the direction given by $\ta = \tan^{-1} [(J_2 - J_1)/\sqrt{3} (J_2 +
J_1)]$, and is maximum in the perpendicular direction. These
statements should be interpreted with some care; $\la \vcr^2
\ra_\ta$ may be small for some value of $\ta$ either due to a
cancellation between positive and negative correlations or because
$\la O_{\vcr}^{\rm 2D} \ra$ is small in that direction.

We note that the sum rule discussed in Sec. \ref{sumrule} is also valid in
2D, and we get $\sum_{\vcr} \la O_{\vcr}^{\rm 2D} \ra = -1 + 2 p_{\vec 0}
= 1$ regardless of how $J_3$ is varied from $- \infty$ to $\infty$.

\subsection{Entropy}
\label{ent}

\begin{figure}
\rotatebox{0}{\includegraphics*[width=\linewidth]{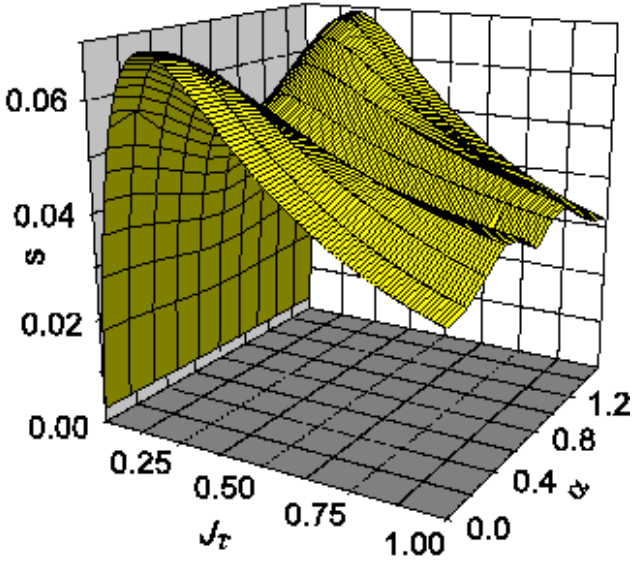}}
\caption{Plot of the entropy density $s$ as a function of $J\tau$
and $\alpha=\tan^{-1} (J_2/J_1)$. The entropy density peaks when
$\left<O_{\vec 0}\right>$ crosses from $-1$ to $1$ as discussed in
the text.} \label{fig6} \end{figure}

As discussed in Sec. \ref{1db}, for each momentum $\vk$, the final
density matrix is effectively diagonal, with entries $1 - p_{\vk}$
and $p_{\vk}$. The density matrix of the entire system takes the
product form $\rho = \bigotimes \rho_{\vk}$. The von Neumann entropy
density corresponding to this state is given by \beq s ~=~ -
~\frac{1}{A} ~\int d^2 \vk ~[~ (1 - p_{\vk}) \ln (1 - p_{\vk}) ~ +~
p_{\vk} \ln p_{\vk} ~], \label{entropy1}\eeq where the integral
again goes half the Brillouin zone. Let us now consider the
dependence of this quantity on the quenching time scale $\tau$
\cite{sen1}. If $\tau$ is very small, the system stays in its
initial state and $p_{\vk}$ will be close to 1 for all values of
$\vk$; for the same reason, $\la O_{\vec 0} \ra$ will remain close
to 1. If $\tau$ is very large, the system makes a transition to the
final ground state for all momentum except near the line described
in Eq. (\ref{line}). Hence $p_{\vk}$ will be close to 0 for all
$\vk$ except near that line, and $\la O_{\vec 0} \ra$ will be close
to -1. In both these cases, the entropy density will be small. We
therefore expect that there will be an intermediate region of values
of $\tau$ in which $s$ will show a maximum and $\la O_{\vec 0} \ra$
will show a crossover from $-1$ to 1. A plot of $s$ and as a
function of $J\tau$ and $\alpha$, shown in Fig.\ \ref{fig6} confirms
this expectation. We find that the entropy reaches a maximum for the
intermediate value of $J\tau$ where $\la O_{\vec 0} \ra$ crosses
over from $-1$ to 1 for all values of $\alpha$.

\section{Extended Kitaev Model}
\label{ekit}

The extended Kitaev model, described by $H_2$ (Eq.\ (\ref{extkit1})),
can also be mapped, using the Majorana transformation given by Eq.\
(\ref{maj2d}), to a Fermionic Hamiltonian
\bea H'_1 &=& i J_4 \sum_{(j,l) \in A} \left[a_{j,l} a_{j+2,l} - b_{j,l+1}
b_{j+2,l+1} \right] ~+~ H_{\rm 2D}, \non \\
& & \label{exkit1} \eea
where $H_{\rm 2D}$ is given by Eq.\ (\ref{kitham2d1}). We note that in
this model, just as for $H_{\rm 2D}$, $D_n$ commutes with all the
terms in the Hamiltonian and the ground state corresponds to $D_n=1$
for all links of the honeycomb lattice. Thus, in momentum space,
$H'_1$ reduces to a bilinear $2$ by $2$ matrix Hamiltonian $H'_2 =
\sum_{\vk} \psi(\vk)^\da H'_3(\vk) \psi(\vk)$, where
\bea H'_3(\vk) &=& 2 \Bigg\{~[J_1 \sin ({\vk} \cdot {\vec M}_1) ~-~ J_2 \sin
({\vk} \cdot {\vec M}_2)] ~\si^1 \non \\
&& + [J_3 ~+~ J_1 \cos ({\vk} \cdot {\vec M}_1) ~+~ J_2 \cos ({\vk} \cdot
{\vec M}_2)] ~\si^2  \non \\
&& - J_4 \sum_k \sin(\sqrt{3}k_x) \si^3 \Bigg\}. \label{exkit2} \eea
This can be diagonalized to obtain the energy eigenvalues
\bea E_{\vk}^{'\,\pm} &=& \pm 2 \Bigg( J_4^2 \sin^2(\sqrt{3} k_x)
+ [J_3 ~+~ J_1 \cos ({\vk} \cdot {\vec M}_1) \non \\
&& + J_2 \cos ({\vk} \cdot {\vec M}_2)]^2 + [J_1 \sin ({\vk} \cdot {\vec M}_1)
\non \\
&& -J_2 \sin ({\vk} \cdot {\vec M}_2)]^2 \Bigg)^{1/2} \label{exkit3} \eea
Note that the presence of a non-zero $J_4$ introduces a gap in the
spectrum (except when $\sqrt{3}k_x = n \pi$) for all values of
$J_1$, $J_2$ and $J_3$. Thus the quench of $J_4$ ($J_4 = J
(t/\tau)$) carries the system through a critical point at $t=0$
provided $|J_1-J_2| \le J_3 \le (J_1+J_2)$.

\begin{figure}
\rotatebox{0}{\includegraphics*[width=\linewidth]{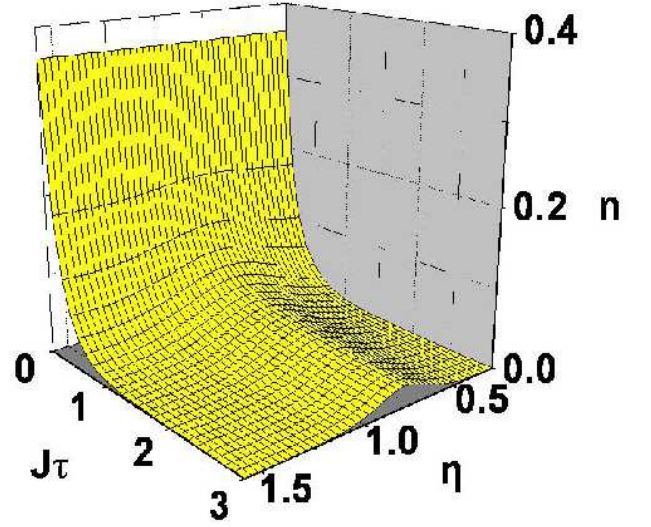}}
\caption{Plot of the defect density as a function of $\eta = J_3/J_1$ and
$J\tau$ for $J_1=J_2=J$.} \label{fig9} \end{figure}

The probability $p_{\vk}$ of defect formation in such a quench, where the
system evolves according to Landau-Zenner dynamics, can be read off from
Eqs.\ (\ref{exkit2}-\ref{exkit3}) as
\bea p_{\vk} &=& e^{-\pi \tau (E_{\vk}^{'\,\pm})^2|_{J_4=0}/| 2 J \sin
(\sqrt{3} k_x) |}. \label{exkit4} \eea
The density of defects is thus given by $n = \int d^2 \vk p_{\vk} /A$,
where the integral is taken over half the Brillouin zone defined by
the triangle with vertices lying at
$(k_x,k_y)=(2\pi/\sqrt{3},0),(0,2\pi/3),(0,-2\pi/3)$ and $A$ is the
area of this region. A plot of the defect density as a function of the
quench rate $\tau$ and $\eta= J_3/J_1$ for $J_1=J_2=J$ is
shown in Fig.\ \ref{fig9}. Note that for large quench time $\tau$,
the maximum contribution to the quench comes from around the
momentum $\vk_0 = (k_{x0},k_{y0})$ for which $E_{\vk_0}^{'\rm
\pm}|_{J_4=0}$ vanishes. Around this point $p_{\vk} \sim \exp[-\pi J \tau
\sum_{\alpha,\beta=x,y} f_{\alpha \beta} (\vk_0) (\vk -\vk_0)_{\alpha} (\vk -
\vk_0)_{\beta}]$ so that $n \sim 1/\tau$ in accordance with the prediction of
the general formula Eq.\ (\ref{gquench1}) for $d=m=2$ and $\nu=z=1$.

Next, we look at the defect correlation functions for the extended
Kitaev model. To this end, we define the operator
\beq O_{\vcr}^{\rm ext} ~=~ i ~\left(a_{\vec n} a_{\vec n+\vcr}-b_{\vec n}
b_{\vec n +\vcr} \right) \label{op1} \eeq
and consider its expectation value for ${\vcr} \ne {\vec 0}$. Here $\vcr =
(\sqrt{3}n_1+\sqrt{3}n_2/2,3n_2/2)$ (with integers $n_1$ and $n_2$) specifies
the sites of the honeycomb lattice. (For ${\vcr} = {\vec 0}$, $O_{\vcr}^{\rm
ext}$ vanishes since $a_{\vec n}^2 = b_{\vec n}^2 =1$).

For $J_4 \to \mp \infty$, the model reduces to a set of decoupled chains
involving Majorana fermions on nearest neighbor sites. For this model, it
is known \cite{shastry} that for $\vcr \ne \vec 0$,
\beq \left< O_{\vcr}^{\rm ext} \right> ~=~ \mp \delta_{n_2,0} ~
\frac{2}{\pi n_1}~ [(-1)^{n_1} ~-~ 1] \eeq
in the ground states for $J_4 \to \mp \infty$ respectively.
For generic values of $J_4$ and for a mixed final state after
the quench characterized by a defect probability $p_{\vk}$, we find
\bea \left< O_{\vcr}^{\rm ext} \right> &=& -\frac{8}{N} ~\sum_{\vk}
\left< a_{\vk}^\da a_{\vk}- b_{\vk}^\da b_{\vk} \right> \sin (\vcr \cdot
{\vk}) \non \\
&=& \delta_{n_2,0} ~\frac{2}{\pi n_1}~ [(-1)^{n_1} ~-~ 1] \non \\
& & +~ \frac{4}{A} \int ~d^2 \vk ~sgn [\sin (\sqrt{3} k_x)] ~p_{\vk} ~\sin
(\vcr \cdot {\vk}). \non \\
& & \label{op2} \eea
The sign of $\sin (\sqrt{3} k_x)$ appears in Eq. (\ref{op2}) because for $J_4
\to \infty$, the ground state of Eq. (\ref{exkit2}) has $\left< a_{\vk}^\da
a_{\vk} - b_{\vk}^\da b_{\vk} \right> = \pm 1$ depending on whether $\sin
(\sqrt{3} k_x) > 0$ or $<0$ respectively.

\begin{figure}
\rotatebox{0}{\includegraphics*[width=\linewidth]{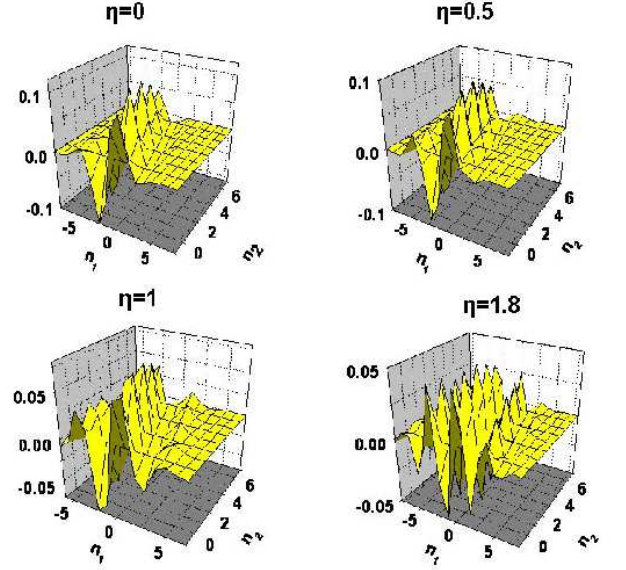}}
\caption{Plot of the defect correlation function (sans the first
term with the delta function peak in Eq.\ (\ref{op2})) with $J\tau=3$
and $J_1=J_2= J$ for several representative values of
$\eta=J_3/J_1$. See text for details.} \label{fig10}
\end{figure}

To obtain an analytical understanding of the nature of the
correlation function, we look at the case where
$J_1=J_2=J$, $J_3= \eta J$ and $J\tau \to \infty$. Note that one
needs the condition $0\le \eta \le 2$ for the system to pass through
a gapless (critical) point during the quench. In this case, the main
contribution to the last term of the correlation function $\left<
O_{\vcr}^{\rm ext} \right>$ in Eq.\ (\ref{op2}) comes from $\vec k=\vec k_0
=((2/\sqrt{3}) \cos^{-1} (-\eta/2),0)$ where $p_{\vk =\vec k_0} =1$.
Thus for $J\tau \to \infty$ one gets
\bea \left< O_{\vcr}^{\rm ext} \right> \simeq \sin\left[ \left\{ 2n_1 +
n_2 \right\} \cos^{-1} \left(\frac{-\eta}{2} \right)\right] \label{op3} \eea
where we have omitted the first term (proportional to $\delta_{n_2 ,0}$) in
Eq. (\ref{op2}). Eq.\ (\ref{op3})
clearly brings out the dependence of the spatial anisotropy of the defect
correlation function as a function of $\eta$. In particular, for $\eta=0$,
$\left< O_{\vcr}^{\rm ext} \right> \sim \sin \{ (n_1+ n_2/2) \pi \}$,
so that its sign alternates between sites with odd and even values of
$n_1$ (if $n_2$ is odd). Similarly, for $\eta=2$, $\left< O_{\vec
r}^{\rm ext} \right> \sim \sin \{ (2n_1 + n_2)\pi \} \sim 0$.
Such a behavior of the correlation function is qualitatively supported by the
numerical computation of $\left< O_{\vcr}^{\rm ext} \right>$ for $J_1=J_2=J$,
$J_3= \eta J$ and $J\tau =3$ as shown in Fig.\ \ref{fig10}. We find that for
$\eta =0$ (top left plot of Fig.\ \ref{fig10}), it alternates between odd and
even $n_1$ sites, while for $\eta$ close to $2$ (bottom right plot in Fig.\
\ref{fig10}), the correlation function is much smaller than for $\eta =0$.

\begin{figure}
\rotatebox{0}{\includegraphics*[width=\linewidth]{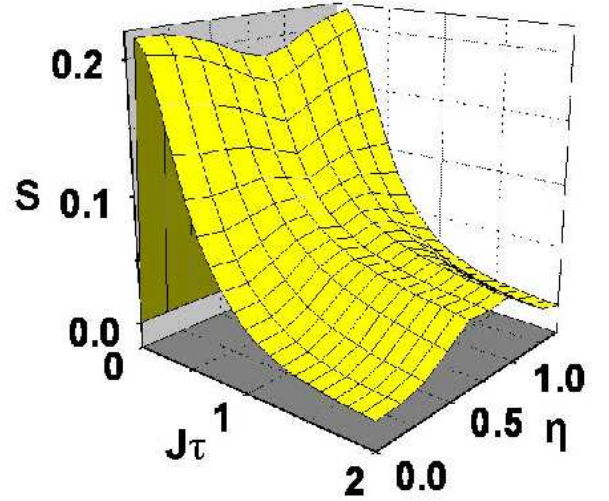}}
\caption{Plot of the entropy density $s$ as a function of quench
time $\tau$ and $\eta = J_3/J_1$.} \label{fig11} \end{figure}

Finally, we compute the entropy generated due to such a quench
process given by Eq.\ (\ref{entropy1}) where $p_{\vk}$ is given by Eq.\
(\ref{exkit4}). A plot of the entropy density as a function of $J\tau$
and $\alpha=\tan^{-1} (J_2/J_1)$ with $J_1=J_3=J$ is shown in Fig.\
\ref{fig11}. Once again we find, similar to that in the Kitaev
model, that the entropy density peaks for intermediate value of $\tau$.

\section{Discussion}
\label{concl}

In conclusion, we have studied the quench dynamics of the Kitaev
model in 1D and 2D and the extended Kitaev model in 2D. For the 1D
Kitaev model and the 2D extended Kitaev model, we have shown that
the defect density scales as $1/{\tau}^{d/2}$ with the quench time
$\tau$, in accordance with the general results of Ref.\
\onlinecite{anatoly1}. For the 2D Kiatev model, where the quench
takes the system through a gapless line, we found that the scaling
of the defect density with $\tau$ changes due to the presence of a
critical {\it line} instead of a critical {\it point}. In this
context, we have presented a general formula for the quench rate
dependence of the defect density for a $d$ dimensional system when
the quench takes such a system through a $d-m$ dimensional critical
surface. We have also computed the defect correlation function for
such quenches by an exact computation of all independent non-zero
spin correlation functions in the defect ground state. In $d=2$, we
have found that the defect correlation function exhibit spatial
anisotropy and studied the dependence of this anisotropy with the
system parameter. Finally, we have computed the entropy generated in
such processes and have shown that the entropy peaks approximately
at values of the quench rate for which the defect correlation
function changes from $-1$ to $1$.

There have been proposals for experimentally realizing the Kitaev model in
systems of ultracold atoms and molecules trapped in optical lattices
\cite{duan}. If this can be done, the evolution of the defect
correlations with various parameters (such as $J_2/J_1$ as shown in Fig.\
\ref{fig7}) can, in principle, be experimentally detected by spatial noise
correlation measurements as pointed out in Ref. \onlinecite{altman}.

Finally, we would like to note that the quench dynamics of the $XXZ$ spin-1/2
chain has been recently studied with the Hamiltonian being varied along a line
in parameter space where the model is critical \cite{pellegrini}. In momentum
space, the model only has a
finite number of critical points, but the system stays close to those
critical points for a long time. This is a different situation from the
one that we have analyzed in Sec. III where there is a line of critical
points in momentum space; hence our results for the scaling of the defect
density are not applicable to the work in Ref. \onlinecite{pellegrini}.

We thank Amit Dutta and Anatoly Polkovnikov for stimulating discussions and
several important suggestions. DS thanks DST, India for financial support
under the project SR/S2/CMP-27/2006.

\end{document}